\documentclass[letterpaper]{article}

\usepackage[T1]{fontenc}

\usepackage{geometry}
\usepackage{setspace}

\usepackage[
  style        = chem-acs,
  articletitle = true,
  doi          = false, 
  isbn         = false  
]{biblatex}
\usepackage{graphicx}
\usepackage{float}
\newfloat{scheme}{htbp}{los}
\floatname{scheme}{Scheme}
\floatname{chart}{Chart}
\newfloat{graph}{htbp}{loh}

\usepackage{chemformula} 
\usepackage[version = 4]{mhchem} 

\usepackage{multicol}

\usepackage{authblk}
\author[1]{Katherine P. Bronstein}
\author[1]{Noah A. Harris}
\author[1]{Aleczander J. Harder}
\author[1]{Jennay L. Edmondson}
\author[1]{Jesse A. Rodr{\'i}guez*}
\affil[1]{Department of Mechanical, Industrial and Manufacturing Engineering, Oregon State University, Corvallis, OR, USA}

\title{\textit{In-Situ} Inverse Design of a Plasma Metamaterial Beam Steering Device}
\date{*Email: jesse.rodriguez@oregonstate.edu}

\begin{document}

\maketitle

\begin{abstract}

  Inverse design is a commonly used methodology for creating devices that manipulate electromagnetic (EM) waves by algorithmically modifying device parameters to achieve a desired functionality. Utilizing plasma, a dynamically tunable medium, allows the optimization of the design process to be conducted directly on the experimental hardware (\textit{in-situ}). A key advantage of this method is the creation of devices that are inherently switchable and dynamically reconfigurable. Bayesian optimization is used to tune the plasma density of 91 independent discharges that make up a plasma metamaterial (PMM) device to steer incoming EM waves to desired exit waveguides. Measurements were conducted in an automated loop where a vector network analyzer records the PMM transmission characteristics for each device setting. By relying only on measured scattering parameters, this gradient-free approach is robust to experimental drift and noise and does not require complex full-wave models. Significant performance improvements over traditional simulation-based (\textit{in-silico}) inverse design are demonstrated, with \textit{in-situ} Bayesian optimization achieving up to 10,000× higher isolation between ports than the best \textit{in-silico} design at the same target frequency. This work also presents guidelines for applying Bayesian optimization to noisy, high-dimensional physical systems.
\end{abstract}

\section*{Keywords}

plasma metamaterial, inverse design, beam steering, reconfigurable photonics, optical computing




\begin{multicols}{2}
\section{Introduction}

Inverse design has become a powerful way to discover electromagnetic devices that would be difficult to design by hand. In this approach, the device permittivity distribution is encoded by a set of tunable parameters, and those parameters are optimized so that the measured or simulated fields match the desired behavior. Plasma metamaterials (PMMs) are a useful platform for inverse design because each plasma element can be tuned electronically over a wide range of effective permittivities, making the same hardware reusable for many different device functions that can be switched dynamically.

In electromagnetics, inverse design is widely used to create devices that are effectively impossible to design by hand \cite{su2020SPINS, miller2013, Liu2013, Hughes2018, Molesky2018, Christiansen2021, Andrade2019}. These systems are constrained by Maxwell’s equations and require numerical solvers to evaluate and optimize each design; heretofore referred to as \textit{in-silico} inverse design. Many inverse-designed electromagnetic devices have been demonstrated, including metalenses \cite{Christiansen2020, Meem2020}, filters and wavelength demultiplexers \cite{Piggott2015}, photonic crystals \cite{Borel2004, Minkov2020, burger2004, Lin2016, Jiang2020}, conformal volumetric metamaterials \cite{Huang2023}, and even optical computing devices like boolean logic gates \cite{Rodriquez_2021} and matrix multiplication devices \cite{Nikkhah2024}. These operations underlie tasks such as signal processing, communication, and machine-learning inference, so faster implementations translate directly into higher-throughput specialized computing hardware. 

Unlike prefabricated metamaterial structures \cite{Nikkhah2024, Pestourie2018, chung2020, Lin2018}, PMMs offer real-time control of wave propagation by tuning plasma density \cite{RodriguezCappelli2022, RodriguezCappelli2023, Sakai2012}. The epsilon near zero (ENZ) regime is crucial to operating these PMM-based devices. By tuning plasma density near the critical frequency, the effective permittivity can approach zero, enabling unique field confinement, phase control, and impedance matching effects \cite{Engheta2013, Liberal2017}. In PMMs, this is achieved through the electrical control of the plasma frequency, making the ENZ behaviors dynamically accessible and providing an advantage over conventional dielectric metamaterials. This allows the PMMs to serve many different purposes without any refabrication. Earlier plasma photonic crystal experiments have demonstrated bandgap devices, waveguides, and lattice-resonance effects in related discharge geometries, establishing plasma as a viable platform for reconfigurable electromagnetic structures \cite{Righetti2018,Wang2016APL,Wang2016Waveguiding,Sakai2013Functional,Sakai2018NegEps,Wang2019Woodpile,Houriez2022,Mehrpour2022Properties,mehrpour2022tunable}.

The tunability of the plasma elements that constitute a PMM arises from the Drude dielectric permittivity of low temperature plasmas. The permittivity (neglecting the ion's response, an appropriate assumption for low temperature plasma) is of the following form:
\begin{equation}
	\varepsilon=1-\frac{\omega_p^2}{\omega^2+i\omega\gamma}
\end{equation}
where $\gamma$ is the collision/damping rate, $\omega_p^2=\frac{n_ee^2}{\varepsilon_0m_e}$ is the plasma frequency squared, $n_e$ is the electron density, $e$ is the electron charge, $m_e$ is the electron mass, and $\varepsilon_0$ is the free-space permittivity. We can see here that in the collisionless limit, the plasma permittivity can be tuned continuously in the interval $(-\infty,1)$ by varying the plasma density $n_e$ for a given operating frequency $\omega$. This is done in practice by supplying more power to the discharge. By choosing an operating frequency that is close to the plasma frequency of the elements, we can easily make drastic changes to their electromagnetic response; \textit{i.e.} to that of a metallic element or a near-zero-index medium. This dispersion relation not only makes the configuration space infinite as opposed to the binarized parameterization schemes most often used in the inverse design of electromagnetic devices, but more importantly, this allows a single array of plasma elements to perform several functions. Moreover, since the plasma elements can be quickly deactivated or tuned, you also can switch the device on and off or to a different function very quickly in $\sim 10$ ms or less.

A major opportunity created by PMMs is the ability to run inverse design directly on the experimental device and link it to traditional \textit{in-silico} design in a closed discovery loop. In this work, we compare \textit{in-silico} inverse design that utilizes a realistic numerical plasma model \cite{RodriguezCappelli2022} with \textit{in-situ} inverse design that optimizes the plasma frequency of the plasma elements using only measured scattering parameters. Related \textit{in-situ} training ideas have recently been demonstrated in photonic neural networks, where the photonic hardware itself is optimized using measured gradients \cite{Pai2023}. The \textit{in-silico} loop provides physics-aware gradient-based updates but depends on assumptions about density profiles, collision rates, and boundary conditions as well as the dimensionality of the simulation, whereas the \textit{in-situ} loop operates on the real hardware and automatically includes experimental imperfections such as collisional damping, electrode geometry, misalignment, and three-dimensional effects. Comparing the optimal configurations from these two loops reveals where the plasma model and experimental plasma frequency mapping fail to match reality, turning the device into a platform for both performance optimization and plasma-model validation.

\begin{figure*}[htbp!]
    \centering
        \includegraphics[width=\textwidth]{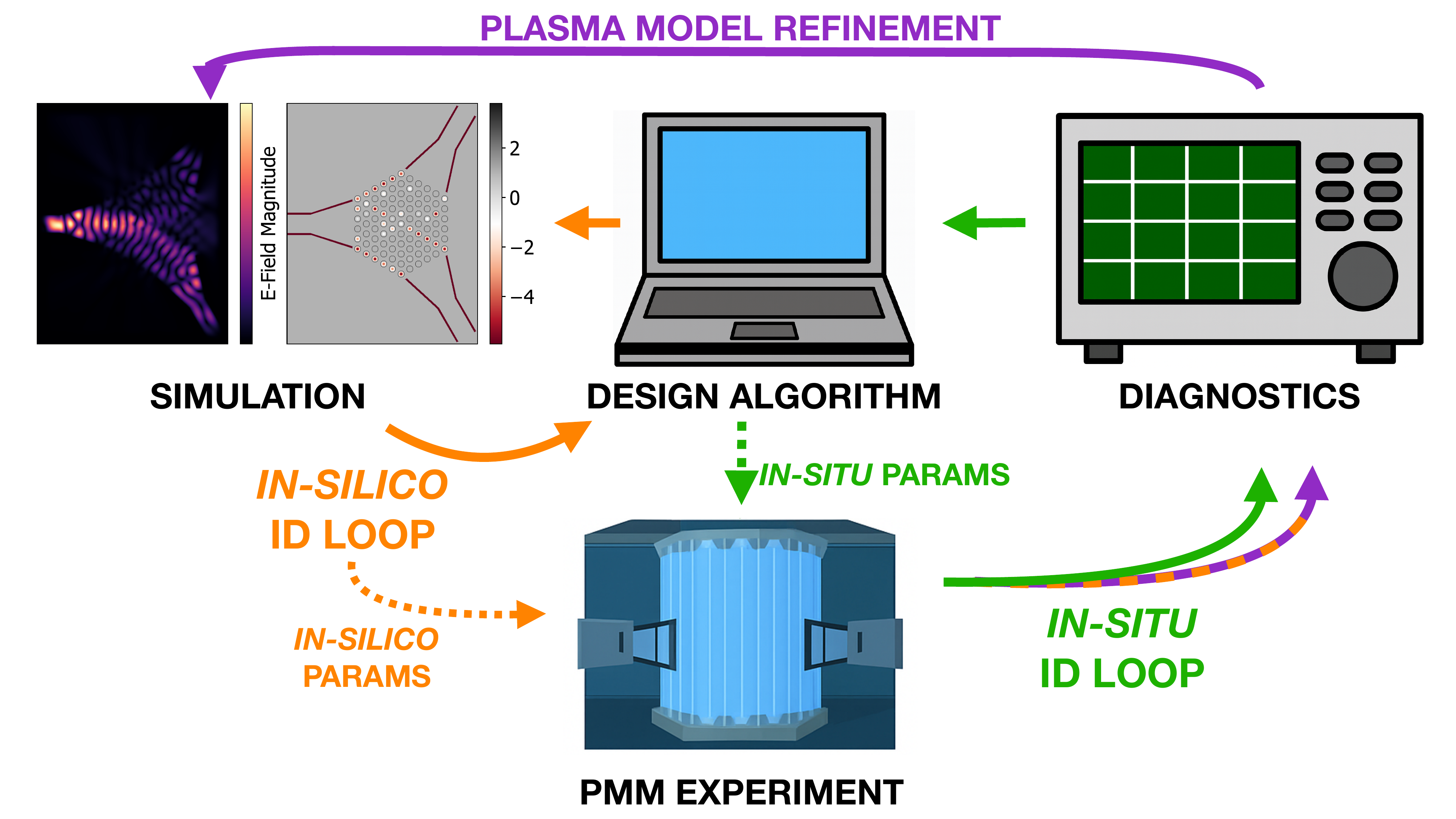}
    \caption{Illustration of the inverse-design workflow showing the complementary roles of \textit{in-silico} and \textit{in-situ} optimization.}
    \label{fig:process}
\end{figure*}

This paired approach turns the PMM into both a reconfigurable device and a probe of its own underlying physics. When the \textit{in-situ} and \textit{in-silico} optimizations disagree, differences in performance and in the resulting plasma-density distribution throughout the device elements highlight specific gaps in our modeling of the plasma elements. Over many optimization runs, the accumulated experimental data can be used to refine the plasma model and to train surrogate models that accelerate future design cycles. Figure~\ref{fig:process} summarizes this workflow: a design algorithm exchanges parameters with both the numerical solver and the physical PMM, while diagnostics close the loop by feeding measurements back into the model. 

In this study, we present the first electromagnetic device which has a domain created entirely via \textit{in-situ} inverse design. We demonstrate how the \textit{in-situ} inverse design process, when coupled with the PMM device paradigm, leads to vastly better performance than \textit{in-silico} inverse design while also enabling devices that are inherently multifunctional and tunable. We begin by describing the PMM hardware, voltage-to–plasma-frequency mapping, and inverse-design objectives and methodology. Next, we compare \textit{in-silico} and \textit{in-situ} beam-steering results for standard, narrowband, and broadband objectives. Finally, we discuss how these results inform plasma-model refinement and future PMM-based electromagnetic wave control, including optical computing devices.

\section{Methods}

\begin{figure*}[htbp!]
    \centering
    \begin{minipage}[t]{0.53\textwidth}
        \centering
        \includegraphics[width=\textwidth]{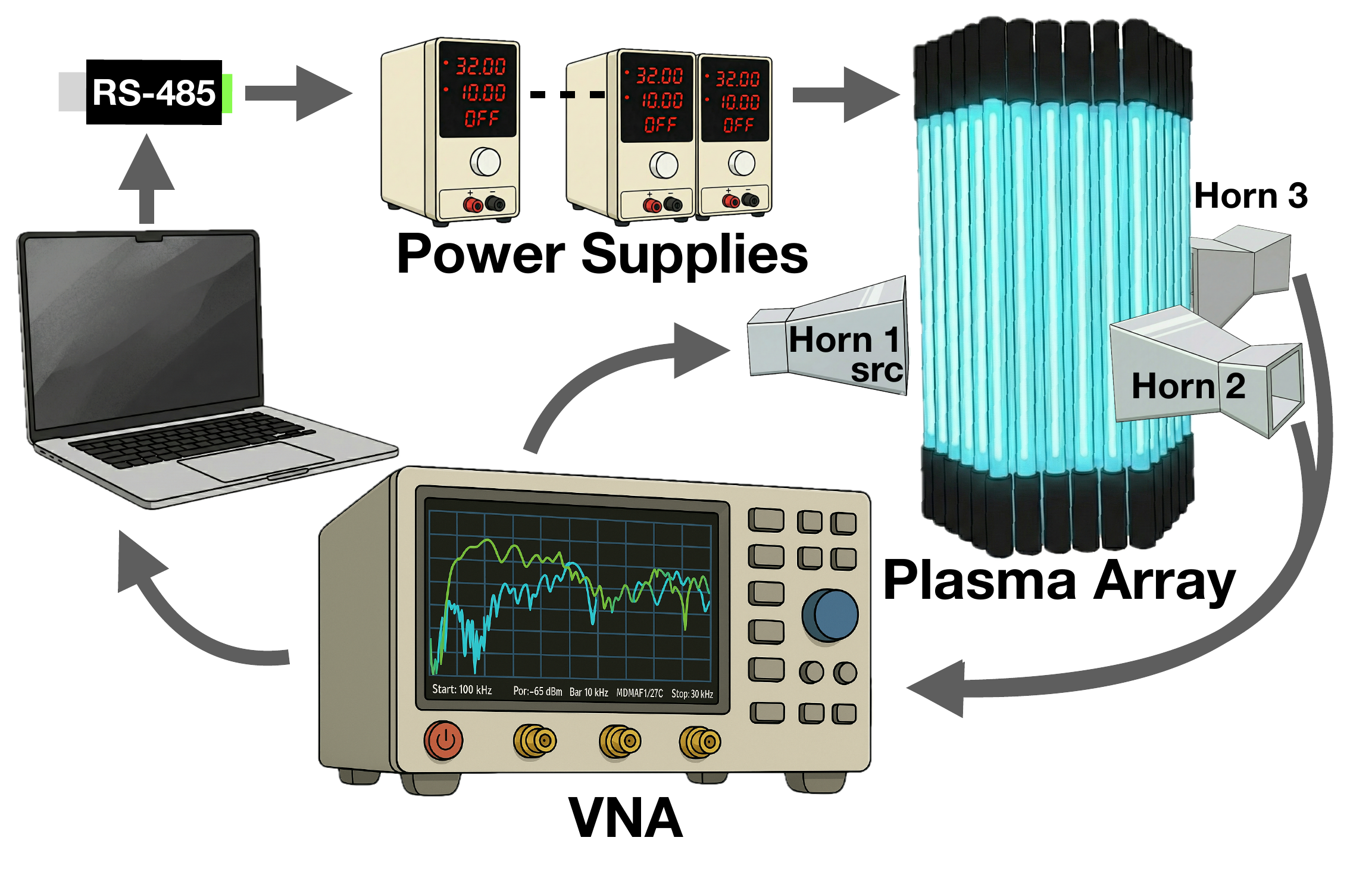}
    \end{minipage}
    \hspace{0.000001\textwidth}
    \begin{minipage}[t]{0.45\textwidth}
        \centering
        \includegraphics[width=\textwidth]{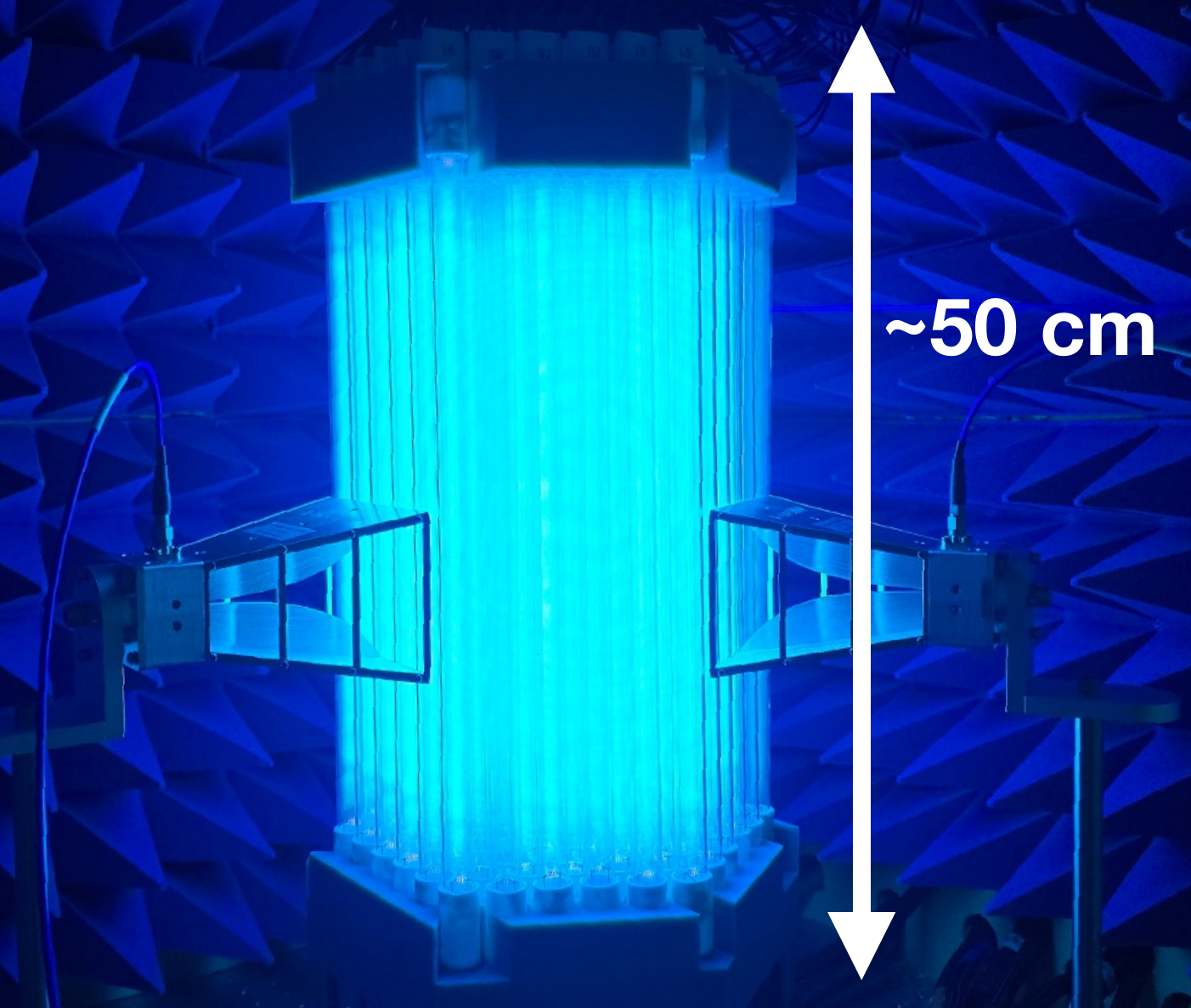}
    \end{minipage}
    \caption{Experimental setup: (left) schematic of the PMM control system, and (right) photograph of the physical implementation of the 91-element plasma array.}
    \label{fig:setup}
\end{figure*}

The device utilized in this study is a two-dimensional plasma metamaterial composed of 91 individually addressable discharge tubes arranged in a triangular lattice in the style of a photonic crystal. The elements form a hexagon with six bulbs along each side, so that the center-to-center spacing (20 mm) is fixed across the array, as seen in Figure \ref{fig:setup}. The spacing is small compared to the 4-10 GHz wavelengths of interest, so the array can behave like a single contiguous metamaterial. The array is mounted between a source microwave horn antenna (port 1) and two receiver horns (ports 2 and 3), and the entire assembly is placed inside a compact anechoic chamber lined with microwave absorber to suppress reflections during \textit{in-situ} optimization and S-parameter measurements, as seen on the right side of Figure \ref{fig:setup}. 

Microwave transmission is measured with a Rohde \& Schwarz ZNB40 vector network analyzer (VNA) configured to measure $S_{21}$ and $S_{31}$ transmission spectra between the input (port 1) and each receiver horn antenna (ports 2 and 3). The VNA is calibrated using an automatic calibration kit (R\&S ZN-Z54) along with the SMA test cables used to connect the VNA to the horn antennas. During optimization of the PMM devices, the VNA is left in a fixed measurement state with a frequency sweep covering the band of interest (0-20 GHz) with 10,000 points and an averaging factor of 10 to reduce noise. For each array configuration selected by the optimizer, the control script triggers a fresh sweep, waits for completion, and then queries the complex $S_{21}$ and $S_{31}$ traces along with the corresponding frequency vector. In post-processing, we convert these to magnitudes in dB and use them to compute scalar objective values.

Each plasma element is a custom UV germicidal discharge tube made of high-purity quartz (relative permittivity $\epsilon =3.8$) that is filled with Argon and a small amount of mercury to provide UV radiation. We used UV tubes instead of commercial lighting bulbs to avoid fluorescent coatings that are opaque at microwave frequencies. The quartz envelope has an outer diameter of 15 mm with a thickness of 1 mm, yielding a 13 mm inner diameter. In operation, the neutral gas temperature is approximately 315-330 K (measured via infrared thermometer), and the discharges are designed for AC drive at 20-50 kHz. The discharge waveform is controlled via DC-to-AC ballast circuits with a maximum driving DC voltage of about 20 V. At this maximum driving voltage, each ballast-discharge circuit draws about 30 W. The discharges have a rise-time of about 14 ms, and a fall-time of about 45 ms. These times are defined according to the plasma's dielectric response; see App. A of ref. \cite{RodriguezThesis2023}.

To enable software-controlled operation, each discharge-ballast circuit is driven by a dedicated programmable Longwei LW-3010 DC power supply. The power supplies are connected in groups of ten to RS-485 communication buses and controlled using MODBUS serial commands from a custom python library. To compare the \textit{in-silico} optimal parameters to those from the \textit{in-situ} optimization algorithm, we introduce a fixed mapping between the simulated element indices and the hardware addresses of the bulbs and power supplies. In the numerical simulations, the design variables form a vector $\rho$ whose entries correspond to elements ordered in a triangular lattice. In hardware, each discharge is linked to that mapping. 

To connect the \textit{in-silico} plasma frequency parameters to experimental control settings, we construct a quasi-experimental mapping from power-supply voltage to plasma frequency. To measure the plasma density, we use a 0-D kinetic solver (BOLSIG+ \cite{Hagelaar2005,SIGLO}) to relate the measured reduced electric field across the discharge $E/n$ to the electron energy distribution and consequently the electron mobility and drift velocity, which can be used to calculate the electron density. For each operating point, we measure the RMS voltage and current across a single discharge. We then estimate the neutral gas density from the fill pressure (200-300 Pa, proprietary to the discharge producer and thus not precisely known) and a range of assumed gas temperatures (315-330 K) via the ideal gas law. Using the discharge length and an estimate of the cathode voltage drop \cite{Hilscher2002}, we infer the effective electric field in the positive column. BOLSIG+ then provides the electron mobility $\mu_e$, which we multiply by the measured electric field to obtain the average electron drift velocity, $u_d = \mu_e E$. Combining this drift velocity with the measured RMS current and the tube cross-sectional area gives an estimate of the electron density $n_e = I_{meas, RMS}/(eu_dA_{tube,inner})$, and thus the plasma frequency. We repeat this procedure over a range of voltage-limited points, sweep over possible fill pressures, temperatures, and cathode voltage drops, and then fit a curve that maps our voltage to plasma frequency, following the approach detailed in Appendix A of Ref. \cite{RodriguezThesis2023}. 

\begin{figure}[H]
    \centering
        \includegraphics[width=\columnwidth]{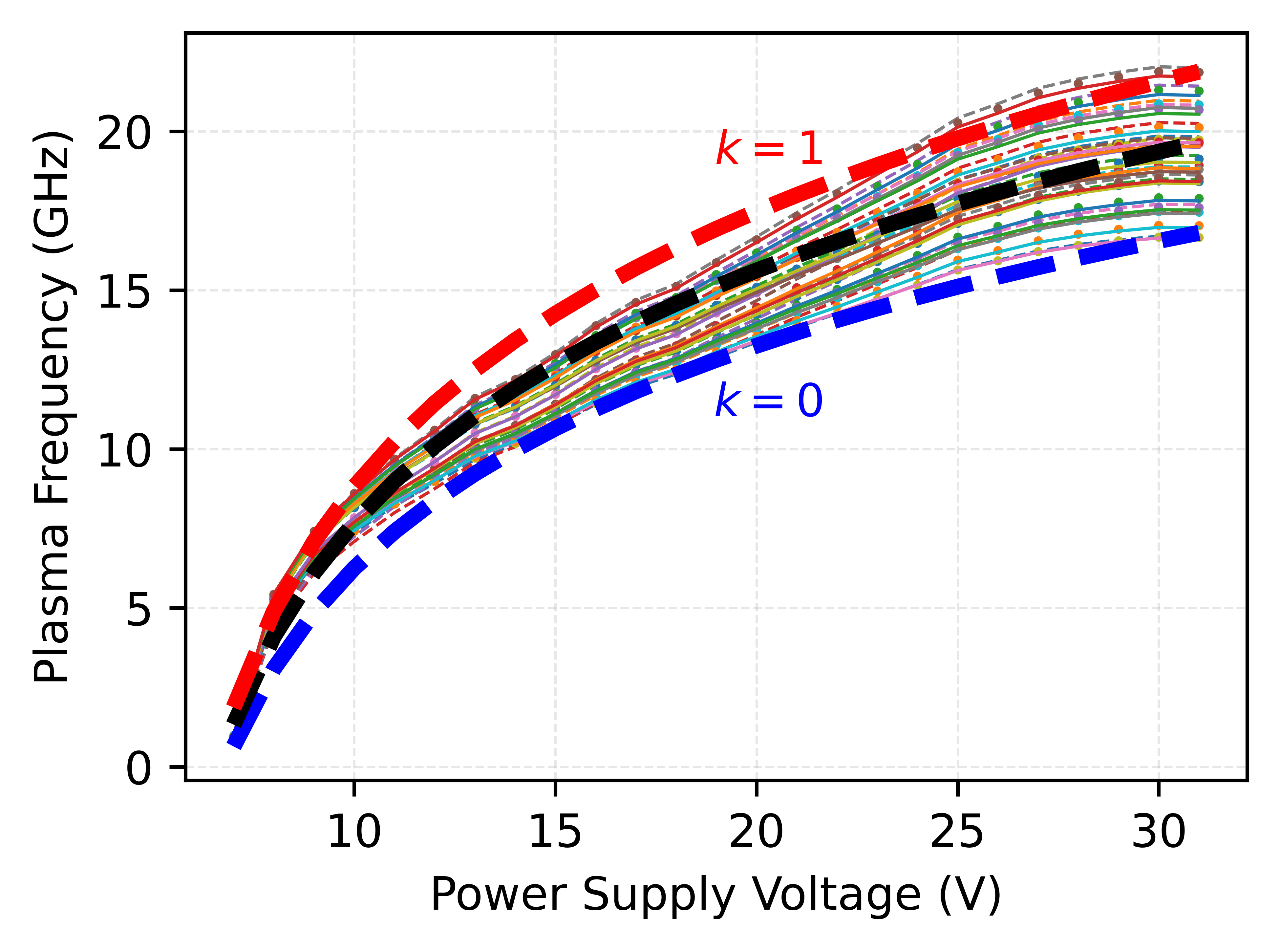}
    \caption{The quasi-experimental mapping for plasma frequency tuning, showing three fitted voltage-frequency mappings (thick dashed curves) that bound the low-, mid-, and high-density cases encoded by $k$ in Equation 2. Each of the thin colored curves plotted underneath the fitted function show the BOLSIG+-based estimate of plasma frequency for a specific combination of gas pressure and temperature (e.g. 220 Pa and 300 K) resulting from a sweep across possible values.}
    \label{fig:mapping}
\end{figure}

The resultant fit for plasma frequency $f_p$ in GHz follows a logarithmic function of the form
\begin{equation}
f_{p,\text{GHz}}=S\left[ (10.5+2.5k)\,\log_{5}\!\left(V - 4.8\right) - 4.5 \right],
\end{equation}

where $k$ tunes between low and high-density envelopes and the global scale factor $S$ shifts the entire curve to account for systematic over- and underestimates of electron density. The mapping is plotted in Figure \ref{fig:mapping}.\

\begin{figure*}[htbp!]
    \centering
        \includegraphics[width=0.89\textwidth]{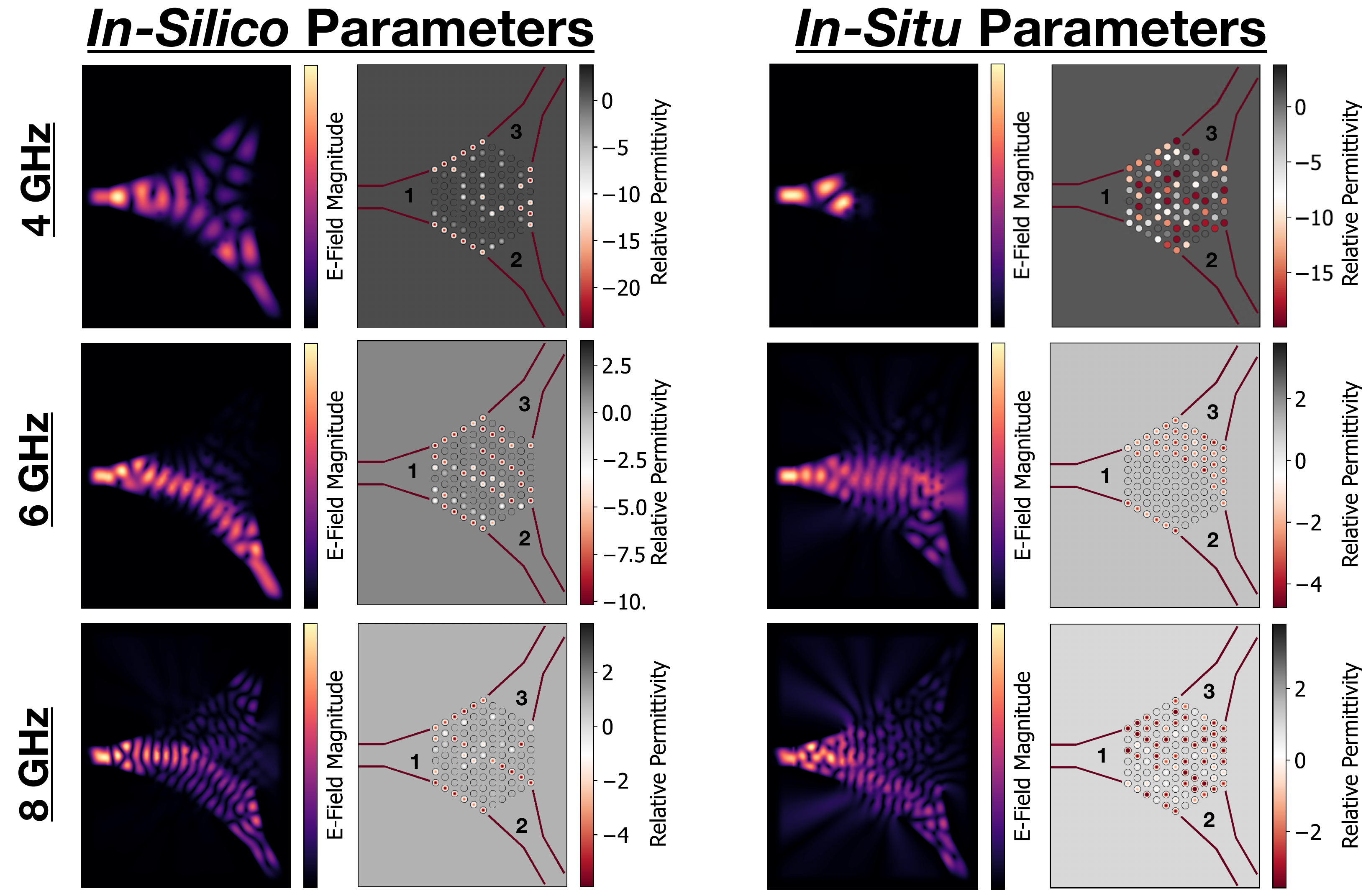}
    \caption{PMM FDFD simulation results at 4, 6, and 8 GHz: steering comparing \textit{in-silico} and \textit{in-situ} parameters, showing stronger steering with the \textit{in-silico} parameters. The left panels for each parameter set plot E-field intensity $|E_z|^2$ in the domain and the right panels show the relative permittivity in the domain. The discharges with lowest permittivity correspond to those with the highest plasma density.}
    \label{fig:sim_results}
\end{figure*}

For the \textit{in-silico} optimization, the device is modeled with an autograd-compliant two-dimensional finite difference frequency domain (FDFD) Maxwell solver, Ceviche \cite{Hughes2019}, where the domain includes the input and output horns, perfectly matched layers at the boundaries, and a hexagonal array of discharges using the plasma element model detailed in \cite{RodriguezCappelli2022}. The source mode $E$-field is polarized along the length of the plasma discharges ($E_z$ with $z$ out of the page in Figure \ref{fig:sim_results}). The design vector $\rho$ encodes the average plasma frequency, and thus permittivity, of each of the elements, and the solver returns the fields for a given source frequency. The inverse design algorithm utilizes objectives based on overlap integrals between the simulated field and the desired waveguide mode, the same as those used in prior work \cite{Rodriquez_2021,RodriguezCappelli2022,RodriguezCappelli2023}. These differentiable objectives are used to optimize $\rho$ using the Adam optimizer \cite{KingmaBa2014} with the learning rate $\alpha$ ranging from 0.01 to 0.05 and the default Adam hyperparameters $\beta_1=0.9$, and $\beta_2=0.999$. It is important to note that since the device is simulated in the frequency domain, the numerical results represent the steady state device behavior for a single frequency, and thus the full-spectrum response of the device is not factored into the \textit{in-silico} optimization routine.

To mitigate any reproducibility issues associated with transient thermal effects or element hysteresis, each optimization follows a standard sequence. Before optimization, the array is allowed to ``warm up" by cycling the entire array on and off at the desired experimental duty cycle for approximately ten minutes. Once the optimization commences, each objective evaluation follows the same procedure: First, the supplies are pulsed to a higher ignition voltage to strike the discharges. Once all bulbs are lit, the voltages are adjusted to their operating values specified by the parameter vector $\rho$ and held for a short time (on the order of a second) to allow any transient effects associated with element tuning to subside. The array remains on while the vector network analyzer (VNA) acquires S-parameters, after which the discharges are switched off and allowed to cool. To avoid overheating the tubes and ballasts, we enforce a 25\% duty cycle (roughly 15 seconds per minute) during extended runs.

To optimize the device \textit{in-situ}, we treat the PMM as a black-box function that maps a set of power-supply settings to a scalar performance metric. The experimental objective functions are defined as discrete sums over differences between the measured scattering parameters at ports 2 and 3. For the standard beam steering objective where port 2 is the desired exit port, we first compute, at each frequency $f$, the dB-scale difference: $d(f) = S_{21,\mathrm{dB}}(f) - S_{31,\mathrm{dB}}(f)$, so that large positive $d$ indicates more transmission into the correct exit port. The objective function then consists of summing $d(f)$ over a specified frequency band $B = [f_{op}-\Delta f/2, f_{op}+\Delta f/2]$ where $f_{op}$ is the desired operating frequency. For the standard objective, $\Delta f = 0.5$ GHz.

\begin{equation}
\mathcal{L}_{\mathrm{standard}} = \sum_{f \in \mathcal{B}} d(f)
\end{equation}

For narrowband beam steering, we desire a narrow band of isolation in favor of the correct output port and minimal difference in transmission between the ports at other frequencies. In this case, we define an in-band window (still with $\Delta f = 0.5$ GHz) around a target frequency $f_0$ and treat everything outside that window as out-of-band (Right and Left). The narrowband objective is computed as

\begin{align}
\mathcal{L}_{\mathrm{narrow}} = w_{\mathrm{in}} \sum_{f \in B} d(f)
  &- w_{\mathrm{oob}} \sum_{f\notin B} d(f)^{2}\notag,
\end{align}

where $w_{\mathrm{in}}=2.0$ and $w_{\mathrm{oob}}=0.5$ are objective hyperparameters used to differentially penalize or promote spectral differences within, below, and above the operating band. These parameters are adjusted by hand, guided by experimental results. The squared penalties ensure that any difference in the spectra out-of-band are discouraged.

The extra-broadband objective is designed to push performance over a wide contiguous frequency range. It rewards isolation across a broad frequency band while penalizing rapid fluctuations of the transmission difference. The set $B$ is expanded to include the entire operating range of the microwave horns, 0-20 GHz. The objective is

\begin{align}
\mathcal{L}_{\mathrm{broadband}} = &\hspace{.1cm}w_{\mathrm{sep}}
  \sum_{f\in B} d(f)\notag\\
  &- w_{\mathrm{flat}}
    \sum_{i} \bigl[d(f_{i+1}) - d(f_i)\bigr]^{2},
\end{align}

where $w_{\mathrm{sep}}=1.0$ and $w_{\mathrm{flat}}=0.05$. The second term penalizes roughness by summing the squared differences between adjacent frequency samples. 

\end{multicols}
\begin{multicols}{2}

We implement Bayesian optimization \cite{Bayes} with a Gaussian-process surrogate (implemented via \texttt{gp\_minimize} from \texttt{scikit-optimize}) to explore each ``black-box" objective. The optimizer proposes a new set of parameters $\rho$ by maximizing an acquisition function (expected improvement) over the bounded design space. For each proposed $\rho$, we apply the simulation to hardware mapping, convert the entries to plasma frequencies and voltage setpoints, execute the ignition and warm-up sequence, measure $S_{21}$ and $S_{31}$, compute the chosen objective (standard, narrowband, broadband), and return it to the optimizer. The initial set of parameter samples used to condition the surrogate are either randomly sampled or a ``warm-start" is conducted using samples from previous runs. The initial parameter vector before the first set of samples corresponds to a configuration with the incorrect exit port effectively ``blocked" by high plasma density elements, so that the most power was directed toward the desired exit at the start of the procedure. Each evaluation is logged along with diagnostic plots of the S-parameters and objective value. To avoid plateaus, the parameters may be manually perturbed to encourage exploration of other regions of the configuration space. We repeat this loop until the objective converges or a predefined evaluation number is reached and then the best performing iteration is selected, yielding \textit{in-situ} optimized configurations for the chosen objective. 
\section{Results}

We first applied \textit{in-situ} optimization to the standard beam steering objective at 4, 6, and 8 GHz. Separately, the device was optimized \textit{in-silico} using Ceviche at the same three frequencies following the procedure laid out in refs. \cite{RodriguezCappelli2022, RodriguezCappelli2023}. We present simulated fields for the optimal parameters from both inverse design approaches (\textit{in-silico} vs. \textit{in-situ}) in Figure \ref{fig:sim_results}. As expected, the \textit{in-silico} parameters produce strong performance in the simulated device, whereas the \textit{in-situ} parameters result in comparatively poor performance. This is in stark contrast to the experimental results on the PMM device, where the trend reverses. 

\begin{figure*}[htbp!]
    \centering
        \includegraphics[width=\textwidth]{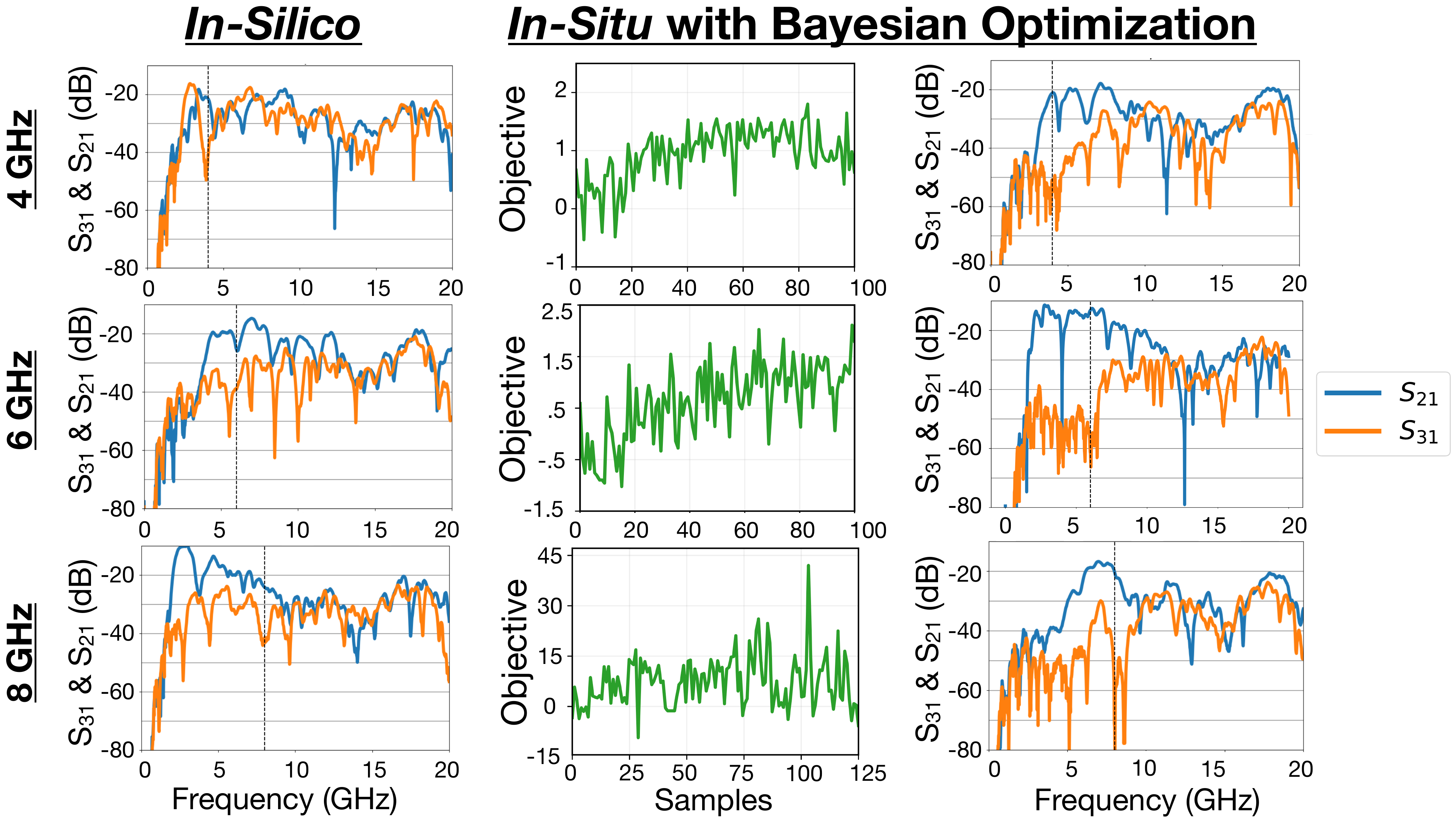}
    \caption{PMM experimental transmission spectra for the standard beam steering objective at 4, 6 and 8 GHz for the \textit{in-silico} parameters (left) and the \textit{in-situ} parameters (right) along with the objective evolution over time (center). The operating frequency is denoted by a vertical dashed black line. The objective values are in arbitrary units. For each target frequency, \textit{in-situ} optimization improves isolation over the \textit{in-silico} parameters by a factor of up to 10$^3$ at 6 GHz and 10$^4$ at 8 GHz on a power scale.}
    \label{fig:standard_results}
\end{figure*}

In Figure \ref{fig:standard_results}, we see that the \textit{in-situ} optimized device produces higher measured isolation between ports 2 and 3 than the \textit{in-silico} design at the target frequencies. At 6 GHz, the \textit{in-silico} design yields

\end{multicols}
\begin{multicols}{2}

\noindent about 15 dB of isolation, while the \textit{in-situ} design achieves about 45 dB isolation, corresponding to a $\sim$1,000x improvement. At 8 GHz, the \textit{in-silico} design likewise gives about 15 dB isolation, and the \textit{in-situ} about 60 dB precisely at 8 GHz, corresponding to about a $\sim$10,000x improvement. Thus, for the same task, the \textit{in-silico} parameters moderately outperform in the numerical model, but the \textit{in-situ} parameters perform far better on the actual device.

\begin{figure}[H]
    \centering
    \includegraphics[width=.9\columnwidth]{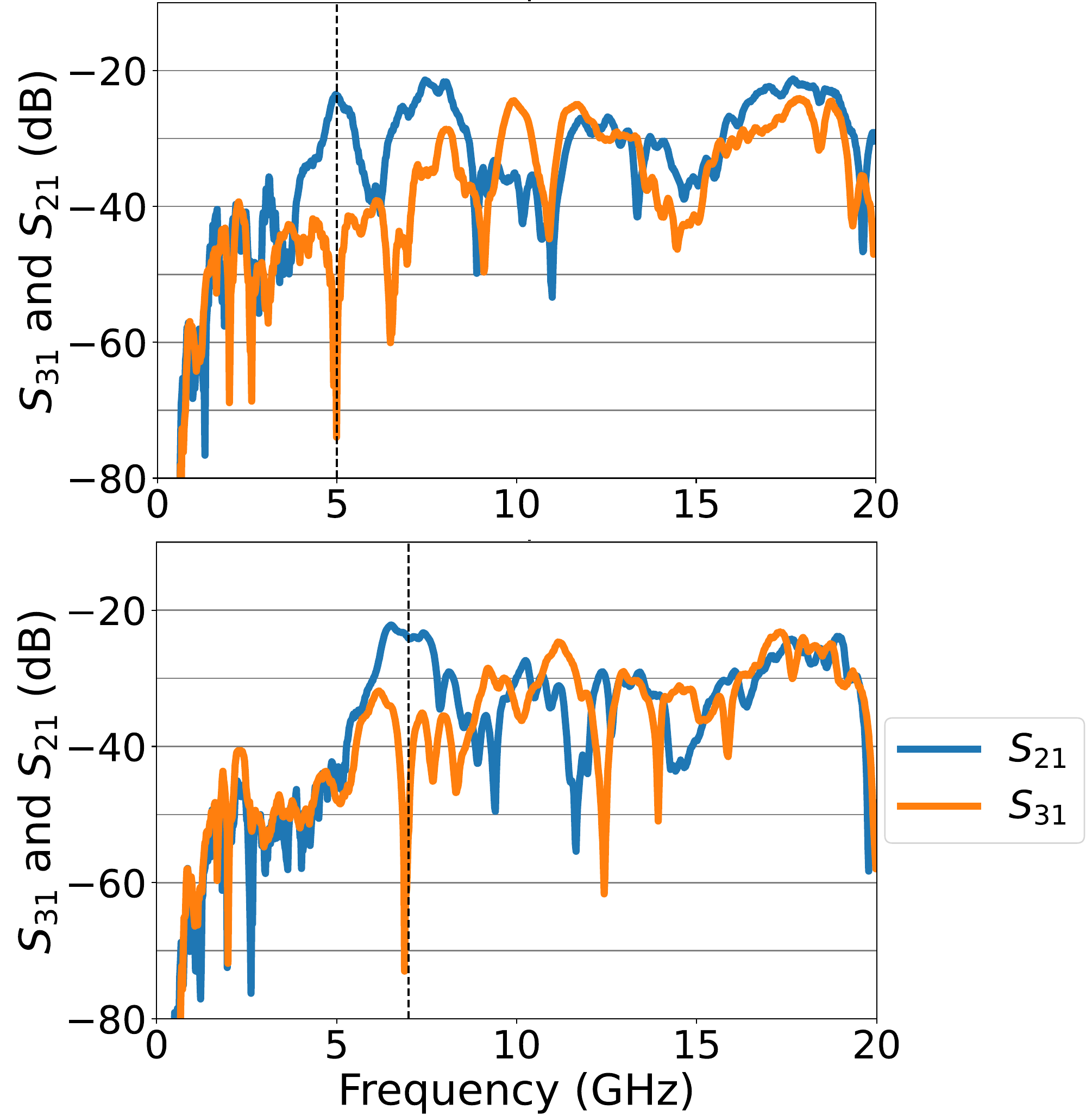}
    \caption{\textit{In-situ} narrow band optimization results targeting 5 (top) and 7 (bottom) GHz. The optimizer finds patterns that yield about 50 dB isolation at the target frequency while maintaining high transmission into the desired port and minimal isolation at other frequencies.}
    \label{fig:narrow_results}
\end{figure}

Next, we applied \textit{in-situ} optimization with the narrowband objective to measure how much isolation the device can provide at a single frequency. For this objective, the initial parameters were set such that all the discharges were inactive. At individual target frequencies between 4 and 8 GHz, including the 4 and 7 GHz cases presented in Figure \ref{fig:narrow_results}, the device achieved approximately 50 dB of isolation between the desired and undesired ports, at least as high as the isolation obtained with the standard objective but over a much smaller bandwidth. Where the standard objective results in isolation from about 2 to 10 GHz of varying strength, the narrow band objective limits the strong isolation to a frequency band of approximately 0.3 GHz in width, with the remainder of the spectrum limited below about 10 dB isolation in either direction.

These experiments were conducted over many weeks and some discharges and ballast circuits were replaced as they degraded or failed. As a check on experimental drift and to determine if these changes had a strong effect, the previously optimized narrowband parameters for 4 and 8 GHz were re-applied after the experimental campaign concluded. The array still produced clear but degraded beam steering with reduced, broadened, or frequency-shifted isolation peaks, indicating that component replacement and long-term drift can partially erode the stored optimum and would likely require re-optimization to fully recover the original narrowband performance. To test this, a follow-up optimization starting from the previously optimal parameters targeting 4 GHz shifted the isolation peak back to the target frequency within three samples ($\sim$30 dB separation) and recovered the full $\sim$50 dB narrowband isolation after just nine samples, showing that deep narrowband performance can be restored quickly. In future implementations, the system could query the power draw of each ballast at a fixed reference voltage during optimization and use changes in that baseline as an indicator of ballast or tube degradation, helping flag drift in the apparatus before it impacts device performance.

Finally, we tested the \textit{in-situ} inverse design procedure on the broadband objective. In this case, the objective rewards a positive dB-scale difference between ports 2 and 3 across as many frequencies as possible while smoothing rapid variations in transmission. As seen in Figure \ref{fig:broad_results}, the optimizer finds a configuration that maintains nonzero isolation over a wide band, with clear separation between ports 2 and 3 from roughly 4 to 12 GHz ranging from 55 dB to 15 dB isolation, with fewer drops in transmission to the correct port across the isolation band, with an additional region of isolation at the higher frequency band from 13 to 15 GHz. 

\section{Discussion}

The results show a clear mismatch between how the plasma metamaterial behaves in simulation and how it behaves in the experiment, highlighting why \textit{in-situ} optimization is necessary. The simulation results in Figure \ref{fig:sim_results} utilize a plasma element model that includes several non-ideal factors, including non-uniform density profiles, collisionality/loss, robustness to random perturbation/error, and quartz envelopes. With

\begin{figure}[H]
\centering
    \includegraphics[width=\columnwidth]{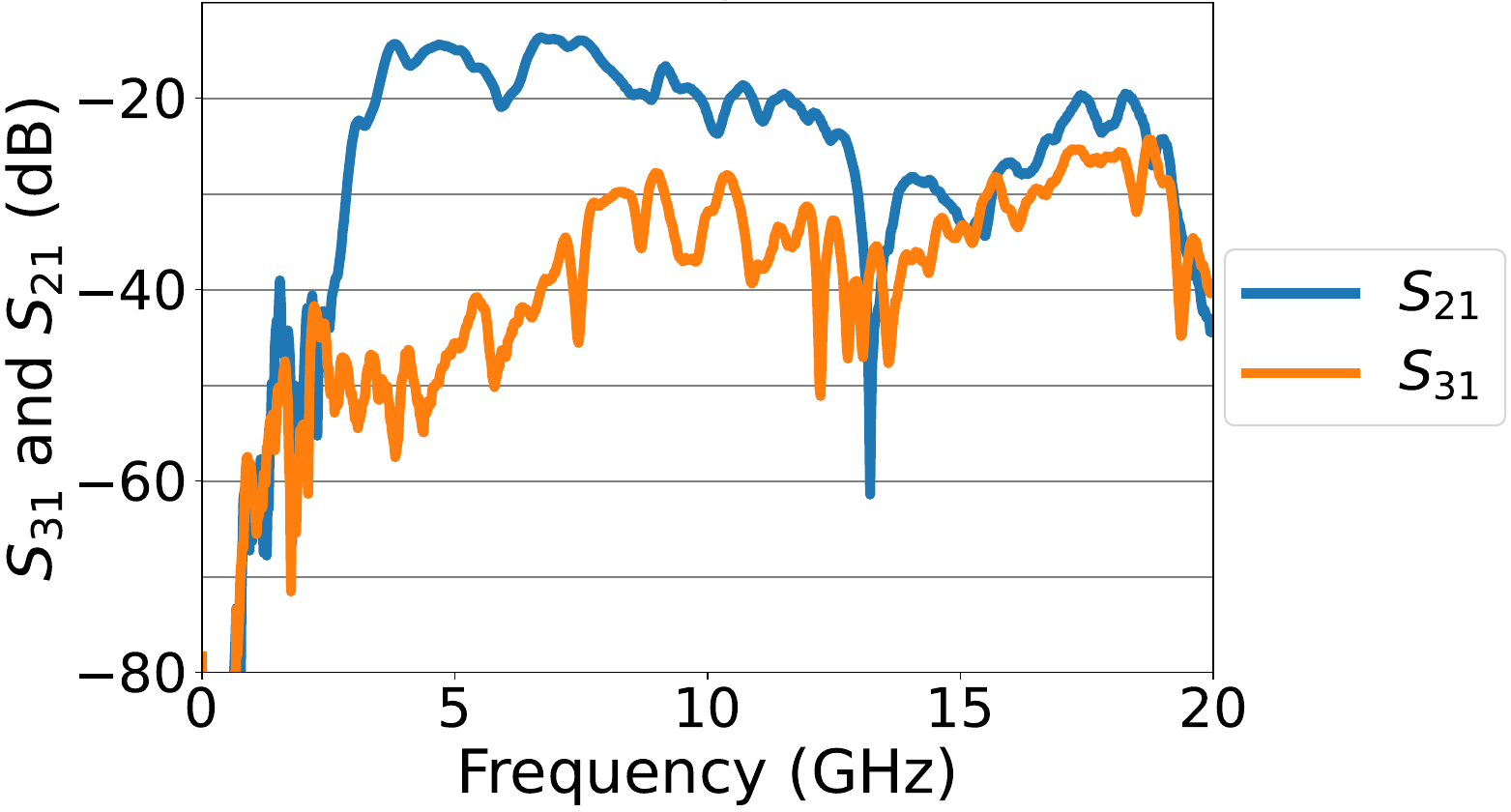}
    \caption{\textit{In-situ} results for the broadband objective. \textit{In-situ} optimization produces a configuration that maintains useful isolation between ports 2 and 3 across a wide frequency range, with clear separation from approximately 4 to 12 GHz.}
    \label{fig:broad_results}
\end{figure}

\noindent these factors considered and the voltage–frequency mapping used in the model with the best case fit tuning factors $k$ and $S$, the numerical model fails to reproduce the stronger performance observed experimentally with the \textit{in-situ} design. This indicates that, although the gradient-based \textit{in-silico} optimization is doing its job and finding a performant design given our plasma model, the model does not fully capture the real plasma and hardware behavior, including the actual density profile, loss, glass response, horn coupling, and perhaps most importantly, three dimensional effects. Although the discharges are much longer than the horn antenna aperture dimensions, the presence of the discharge electrodes or even axial plasma density variations may be playing a role in the shortcomings of the numerical model. At the same time, the \textit{in-situ} pattern still often provides reasonable beam steering performance in the simulations, which suggests that the numerical model is at least capturing the first-order physical behavior of the device. 

The experimental results in Figure \ref{fig:standard_results} show how \textit{in-situ} optimization can side-step errors in numerical models. This objective focused on target frequencies in the range of 4-8 GHz. This range was chosen because the nominal operating plasma frequency (driving at the suggested ballast voltage) of the discharges is estimated to be about 10 GHz, so the device is expected to have the most dynamic range in a frequency band slightly below that value following the Drude permittivity. Within this band, the \textit{in-situ}-optimized patterns provide much higher measured isolation than the transferred \textit{in-silico} design: at 6 GHz the isolation improves by roughly three orders of magnitude, and at 8 GHz it improves by around four orders of magnitude, bringing transmission in port 3 all the way to the noise floor of the VNA. This is because all non-ideal factors are automatically taken into account when optimizing \textit{in-situ}. 

This does, however, present a unique potential method for plasma model refinement. By running both \textit{in-silico} and \textit{in-situ} inverse design in parallel and comparing results, we can comprehensively evaluate the accuracy of our plasma model. As opposed to a slow experiment-simulation feedback loop for model refinement that may neglect edge cases and device performance, the optimization algorithms will probe the full dynamic range of the real and simulated devices, ensuring that if the performance of the optimal configurations from both methods converge, the plasma model must represent the underlying physics extremely well. The large amount of experimental data can also be used to produce low-cost machine learning surrogate models that may speed up both the \textit{in-silico} and \textit{in-situ} inverse design loops.

The narrow band \textit{in-situ} results in Figure \ref{fig:narrow_results} show what the device can do when it is only asked to perform at a single frequency. This resulted in similarly high isolation to the standard objective but in very precise frequency bands. This highly distinct device behavior exhibits the dynamic range of the PMM device. In a single device with $\sim$10 ms switching times, operating frequency bands can be shifted, expanded, and contracted. 

The broadband results in Figure \ref{fig:broad_results} test how wide of a frequency band a single device setting can maintain useful beam steering. The optimized configuration provides clear and consistent separation between ports, but with a lower peak isolation than the narrow band designs. The key difference between this result and those from the standard objective is the influence of the smoothing term. While the standard objective achieves isolation in the correct direction over a similar frequency band in some cases, the strength of isolation is highly variable, sometimes even reversing, such as what is seen in the 6 GHz case. The broadband result also highlights the limited range due to the voltage-plasma-frequency tuning range of the current hardware. Since this objective sought to increase isolation across the entire measured spectrum, the poor performance at higher frequencies is indicative of a shortcoming of the device. This was expected as the maximum plasma frequency of the elements is estimated to be around 15 GHz. Following the Drude permittivity, the plasma elements will become increasingly more transparent above that frequency, reducing their capacity to affect EM wave propagation. This suggests that future devices will need a higher-voltage drive and more resilient discharges to access broader frequency spans. If operated in short pulses, much higher plasma frequencies can be reached without damaging the hardware \cite{chiang2020microplasmas}.

These experiments also expose several limitations of the current \textit{in-situ} optimization loop. The control space is very large (91 plasma discharges), which is not ideal for Bayesian optimization. Because the acquisition function is designed to balance exploration and exploitation, the optimizer will often jump to quite different configurations to ``explore" uncertain regions, then move back towards settings that previously looked good. As a result, the sequence of objective values are non-monotonic and can be very noisy, even when the algorithm is slowly converging over time. In addition, long ignition and warm-up times, duty-cycle limits, and occasional misfires in bulb ignition restrict how many evaluations we can afford to run and introduce slow drifts in the underlying objective, which further challenge the optimizer. Even with these limitations, we nevertheless demonstrate a functioning \textit{in-situ} optimization loop, presenting a great opportunity for improved performance on more complex opbjectives in future work by utilizing better optimization algorithms and workflows.

\section{Conclusion \& Future Work}

The experiments show that \textit{in-situ} optimization enables strong beam steering on the plasma metamaterial device when the plasma settings are tuned directly from measurements, rather than from inverse design algorithms based on simulations. In numerical simulations, \textit{in-silico} inverse design parameters appear to give the best steering and isolation. On the real device, however, those same configurations do not perform as well as those found by Bayesian optimization. This mismatch highlights the impact of modeling uncertainty and further justifies the \textit{in-situ} inverse design approach. We have shown that Bayesian optimization can tune a plasma metamaterial beam-steering device directly from experimental measurements. Crucially, this single device is shown to be rapidly reconfigurable to arbitrary operating frequencies with a very highly variable operating bandwidth, functionality that is not possible with traditional electromagnetic devices. This charts a path towards a fully programmable and generalizable electromagnetic device based on the PMM paradigm.

Moving to fully \textit{in-situ} optimization also removes some of the limitations on device geometry and physics. 3D geometry is generally far too computationally costly to use for \textit{in-silico} inverse design, but it incurs no extra cost in \textit{in-situ} inverse design. Thus, we could use device structures like that of refs. \cite{Wang2019Woodpile,wang2021reconfigurable}. Even for the 2D configuration in this study, each iteration in the optimization procedure can take 5-10 minutes with a 64-core workstation, while the \textit{in-situ} iterations can be limited to mere seconds in practice regardless of the geometry. To combat the loss inherent to gaseous plasma, low-loss dielectric scaffolds could be incorporated into the inverse design workflow via a bi-level optimization scheme \cite{lou2023inverse}. We have also shown in prior work that our plasma sources have a pronounced gyrotropic response when magnetized \cite{Houriez2022, mehrpour2022tunable}, but materials with anisotropic permittivity tensors like magnetized plasma can't be modeled with our simulation tool. By using a large Helmholtz configuration, entire PMM devices can be magnetized to take advantage of the very rich physics inherent to magnetized plasmas, enhancing the dynamic range of our device even further, introducing non-reciprocity, and increasing the potential for other exotic device behaviors like dynamic cloaking \cite{RodriguezThesis2023, rodriguez2020dual}. More sophisticated black box optimization techniques that handle many parameter systems more effectively such as deep learning-based reinforcement learning \cite{haarnoja2018soft} or CMA-ES \cite{nomura2024cmaes} may be used to fully leverage these more complex device phenomena.

\section{Data Availability}
Data available upon reasonable request.

\section{Acknowledgments}
The authors acknowledge the School of Mechanical, Industrial, and Manufacturing Engineering at Oregon State University as well as the Nesbitt Faculty Scholar in Energy Engineering Fund for providing funding to support this work.

\section{Usage of Generative AI}
The authors used generative AI for some limited coding tasks as well as generation of illustrative visual assets in Figures 1 and 2.

\printbibliography
\end{multicols}

\end{document}